\documentclass[runningheads]{llncs}
\usepackage{graphicx}
\usepackage{physics}

\begin{document}
\title{Exponential random graph models for the Japanese bipartite network of banks and firms\thanks{This research was supported by MEXT as Exploratory Challenges on
Post-K computer (Studies of Multi-level Spatiotemporal Simulation of
Socioeconomic Phenomena, Macroeconomic Simulations).
This research used computational resources of the K computer provided
by the RIKEN  Center for Computational Science through the
HPCI System Research project (Project ID: hp180177).
}}
\titlerunning{Exponential random graph models...}
%
\author{Abhijit Chakraborty$^*$ \and
Hazem Krichene  \and
Hiroyasu Inoue \and
Yoshi Fujiwara
}
\authorrunning{A. Chakraborty \em{et al.}}
%
\institute{Graduate School of Simulation Studies, The University of Hyogo, Kobe, Japan
\email{$^*$abhiphyiitg@gmail.com}\\
}
\maketitle              
\begin{abstract}
We use the exponential random graph models to understand the network structure and its generative process
for the Japanese bipartite network of banks and firms. 
One of the well known and simple model of exponential random graph is the Bernoulli model which shows
the links in the bank-firm network are not independent from each other. Another popular exponential random graph model,
the two star model, indicates that the bank-firms are in a state where macroscopic variables of the system can show large fluctuations. 
Moreover, the presence of high fluctuations reflect a fragile nature of the bank-firm network. 

\end{abstract}
\section{Introduction}
Models of networks are useful in studying their structural properties  as well as their dynamical behaviours. 
The approaches to construct models of networks can be classified into two broad categories considering the analogy 
with the theories of gases in statistical physics~\cite{park2004statistical}.  The two approaches are known as 
the kinetic theory approach and the ensemble approach. In kinetic theory approach one considers the possible
mechanisms to replicate some structural properties of the real-world network. For example, the well known 
Barab{\'a}si-Albert model~\cite{barabasi1999emergence} considers preferential attachment mechanisms to construct
a growing network with a fat tail degree distribution. These models are easy to understand and give a qualitative
understanding of the network, but have limitation in quantitative accurate predictions. Thus, these models
do not provide an over-all understanding of the network, rather only mimics few features of the networks. 

The other class of models, the ensemble models,  are based on rigorous probabilistic arguments with a  
solid statistical foundation, useful for accurate predictions and quantitative study of the network. These
models are based on the concept of statistical ensemble implying a large collection of all possible
realizations of the network at particular values of the macroscopic observables. A particular graph in 
the ensemble of networks appears with a probability $P(G) \propto \exp[H(G)]$, where $H(G)$ is known as the network
Hamiltonian. As the probability is an exponential function of the network Hamiltonian, these models are popularly
known as ``exponential random graph (ERG) models". 

The ERG model was first introduced by Holland and Leinhardt~\cite{holland1981exponential}, based on the framework
laid by Besag~\cite{besag1974spatial}. Since the introduction of the ERG models, a variety of network Hamiltonians
have been studied, which include models of random network~\cite{park2004statistical}, reciprocity model of directed
network~\cite{park2004statistical}, the two star model of network~\cite{park2004solution,annibale2015two}, and the Strauss 
model of network with clustering~\cite{strauss1986general,park2005solution}. Far more complex Hamiltonians that include endogenous as well as 
exogenous observables of the network, has also been studied in the social network literature~\cite{lusher2013exponential,wong2015board,simpson2011exponential}.
Moreover, there are many tools such as ERGM~\cite{hunter2008ergm} and SIENA~\cite{ripley2011manual} packages to fit ERG model with social data. 
The problem with the complex non-linear Hamiltonian is that it cannot be solved exactly, the only linear Hamiltonian model can be
solved exactly in the large system size limit. For a non-linear Hamiltonian, it can be solved approximately either using mean-field theory
and perturbation theory or by numerical simulation. 

The ERG model has been studied extensively for monopartite networks except few studies in case of bipartite network~\cite{wang2013exponential}.  
In this paper, our focus is on the Japanese bipartite network of banks and firms. We model the bipartite
network using exponential random graph theory. We study the well known Bernoulli model and two star model to get a deep understanding of the 
network structure of the Japanese bipartite network of banks and firms.

\section{Data}
We use the Nikkei data set for the banks-firms lending-borrowing links in Japan. Lending data are 
available only for the listed firms and are restricted in our work to the long-term loans during 2005. 
Each node in this bipartite network (firms and banks) has its financial statements. However, only listed banks
have available financial statements. Therefore, we consider the unweighted and undirected simple bipartite network for the 
long-term lending-borrowing links between listed firms and listed banks during 2005. The network is formed 
by $M = 127$ banks, $N = 2, 198$ firms and $L = 11, 842$ unweighted long-term links.

\section{Method}
\label{Method}
\subsection{Exponential random graph model}

ERG model is a tie-based statistical model for understanding how network topology emerges by 
estimating how ties are patterned (see~\cite{lusher2013exponential}). Let $X = [x_{ij}]$ be the 
adjacency matrix of an unweighted bipartite network. ERG model is the regression of $X$ with a set 
of endogenous attributes $z_a$ and exogenous attributes $z_e$. $z_a$ represents the network 
statistics configuration, for example, the number of edges or the number of stars. 
$z_e$ represents 
the counts of the node attributes, for example, in case of bank-firm network, the number of links weighted by the profit of 
the firm or the bank. The canonical form of ERG model is given by the following:

\begin{equation}
\mathrm{Pr}_{\Theta}(X = x) = \frac{1}{\kappa(\Theta)}\mathrm{exp}\left(\sum_{a}{\theta_a\cdot z_a(x)} + \sum_{e}{\theta_e\cdot z_e(x)}\right).
\label{eq1}
\end{equation} 

$\kappa$ is a normalizing constant that ensures a proper distribution. Normalization is performed by all possible network realizations, as follows:

\begin{equation}
\kappa(\Theta) \equiv \sum_{y \in X}{\mathrm{exp}\left(\sum_{a}{\theta_a\cdot z_a(y)} + \sum_{e}{\theta_e\cdot z_e(y)}\right)}.
\label{eq2}
\end{equation} 

\subsection{Markov chain Monte Carlo (MCMC) sampling algorithm}
 
Let $x_{obs}$ be the observed graph. We would like to solve the moment equation $\mathrm{E}_{\theta}(z(X)) - z(x_{obs}) = 0$, where $X$ represents the networks sampled with MCMC.

MCMC sampling is used to estimate network statistics $\mathrm{E}_{\theta}(z(X))$. The most commonly used MCMC sampler is the Metropolis-Hasting algorithm, which was
introduced in~\cite{hastings1970monte}.

The MCMC sampler consists of randomly selecting one dyad, one null dyad ($x_{ij} = 0$) or one nonnull dyad ($x_{ij} = 1$). Then, with the Hasting probability $\mathrm{P}(x \rightarrow x')$\footnote{$x$ and $x'$ are network states at simulation steps t and t+1, respectively.}, the state of the dyad is changed (add a link for null dyad or delete a link for nonnull dyad). The Hasting probability is given by the following:

\begin{equation}
\mathrm{P}(x \rightarrow x') = \mathrm{min}\left\{1, \frac{\mathrm{Pr'}_{\Theta}(X = x')}{\mathrm{Pr'}_{\Theta}(X = x)}\right\}.
\label{eq3}
\end{equation}  

\subsection{Stochastic approximation: the Robins-Monro algorithm}

Snijders proposed a stochastic approximation based on the Robins-Monro algorithm to obtain the maximum likelihood estimation (MLE) for the ERG model~\cite{snijders2002markov}. Following~\cite{lusher2013exponential}, this approach is robust and does not require any particular starting point. The stochastic approximation algorithm is based on three phases as described in the following.

\subsubsection{Initialization phase}

With the initial parameter $\tilde{\theta}$, this phase determines the scaling matrix $D_0$. Let $z_{\tilde{\theta}}(x_1), z_{\tilde{\theta}}(x_2),...,z_{\tilde{\theta}}(x_{M_i})$ be the statistics related to networks $x_1, x_2,...,x_{M_i}$ generated with the MCMC sampler based on $\tilde{\theta}$. Let $\mathrm{E}_{\tilde{\theta}}$ be the expectation vector of the network statistics, and let $D$ be the covariance matrix. The scaling matrix is defined as $D_0 = \mathrm{diag}(D)$, and $\theta$ is initialized for the second phase, as follows: $\theta_0 = \tilde{\theta} - a \cdot D_0^{-1} \cdot (\mathrm{E}_{\tilde{\theta}} - z(x_{obs}))$. $a$ is defined as the gain factor, which controls the size of the updating steps ($a = 0.1$ at initialization).

\subsubsection{Optimization phase}

The goal is to solve the moment equation $\mathrm{E}_{\theta}(z(X)) - z(x_{obs}) = 0$ based on the Newton-Raphson minimization scheme. The goal is then to update $\theta$ under different sub phases, where each sub phase $r$ reduces the gain factor $a_r$.

Each sub phase $r$ contains $m$ simulation steps. At simulation step $m+1$, a network is sampled based on the MCMC sampler with $\theta_m$. The update process is defined as follows:

\begin{equation}
\theta_{m+1} = \theta_m - a_r \cdot D_0^{-1} \cdot (z(x_{MCMC}) - z(x_{obs})).
\label{eq5}
\end{equation}

At the end of sub phase $r$, the gain factor is updated, $a_{r+1} = a_r/2$. This optimization procedure is iterated until convergence occurs.

\subsubsection{Convergence phase} We want to check whether the returned value $\hat{\theta}$ from the optimization phase is close to the true MLE. Therefore, $M_c$ networks are sampled based on the MCMC sampler with a value of $\hat{\theta}$. The convergence condition is reached when

\begin{equation}
-0.1 \leq \frac{E_{\hat{\theta}} - z_{obs}}{SD_{\hat{\theta}}} \leq 0.1 ,
\label{eq6}
\end{equation}

where $SD_{\hat{\theta}}$ is the standard deviation of the statistics for the sampled networks.

\section{Results}

\subsection{Bernoulli model of a bipartite network}
In the early 1950s, Solomonoff and Rapoport introduced the first well known model of network, random graph or Bernoulli model of network~\cite{solomonoff1951connectivity}, 
that was later famously studied by Erd\H{o}s and R\'enyi~\cite{erdos1960evolution}. 
This is the simplest model of the network and the analytic solution for the monopartite network using 
exponential random graph technique is shown in~~\cite{park2004statistical}. Here we extend the study for a bipartite network.  
In the Bernoulli model of a bipartite network, links are formed independent of each other and the expected number of links $\langle E \rangle$ is 
the only known observable. The Hamiltonian for this model can be written as $H(G)=\theta E(G)$, where
$\theta$ is the associated parameter with the number of links or it can be thought as inverse temperature
using the analogy with equilibrium statistical mechanics. Using the above expression of the network Hamiltonian,
the probability that the graph $\cal G$ is in state $G$ can be written as $$P( {\cal G} = G) = \frac{e^{\theta E(G)}}{Z}$$
where, the normalization constant $Z = \sum_{\cal G} e^{\theta E(G)}$ is known as the partition function. 
\begin{figure}
\includegraphics[width=\textwidth]{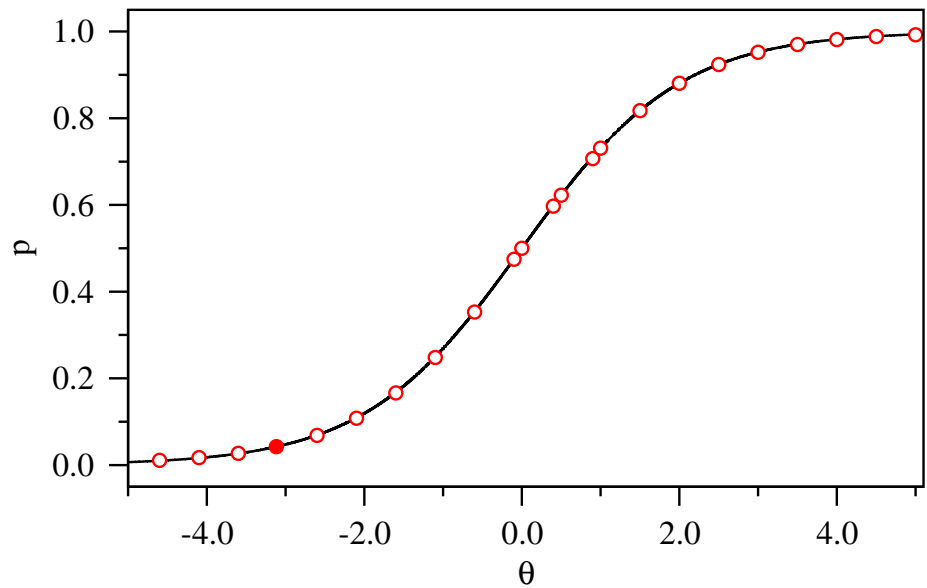}
\caption{The variation of connectance $p$ of the Bernoulli model is plotted as a function of $\theta$ for 
the Japanese bipartite network of banks and firms. 
The solid line represents the exact solution and red circles are the Monte Carlo simulation results.
The filled red circle indicates the simulation result for the observed snapshot of the 
Japanese bipartite network of banks and firms.
} \label{fig1}
\end{figure}

A bipartite network consisting of two distinct node set $\cal{N, M}$ can be represented by a rectangular adjacency matrix with the elements $A_{ij} = 1 \quad
\{1 \le i \le N; 1 \le j \le M\}$ if and only if the $i$-th node of one node set is connected to the $j$-th node of 
the other set and $A_{ij} = 0$ otherwise. The total number of links of the bipartite network 
$E(G) = \sum_{i=1}^{N}  \sum_{j=1}^{M} A_{ij}$. 

Now, we can calculate the partition function as follows:
$$ Z = \sum\limits_{\{A_{ij}\}} e^{\theta \sum\limits_{i=1}^{N}  \sum\limits_{j=1}^{M} A_{ij}} = 
\prod\limits_{i=1}^{N} \prod\limits_{j=1}^{M} \sum\limits_{A_{ij}=0}^{1} A_{ij} = 
\prod\limits_{i=1}^{N} \prod\limits_{j=1}^{M} (1+e^\theta) = (1+e^\theta)^{NM}$$

From the partition function, we can calculate all the network observables:

The free energy of the network $F=\ln{Z} = NM \ln(1 + e^\theta)$

The expected number of edges $$\langle E \rangle = \pdv{F}{\theta} = NM\frac{e^{\theta}}{(1+e^{\theta})}$$

This gives 

$$\theta=\ln[\frac{\langle E \rangle}{(NM - \langle E \rangle)}]$$

\begin{figure}
\includegraphics[width=\textwidth]{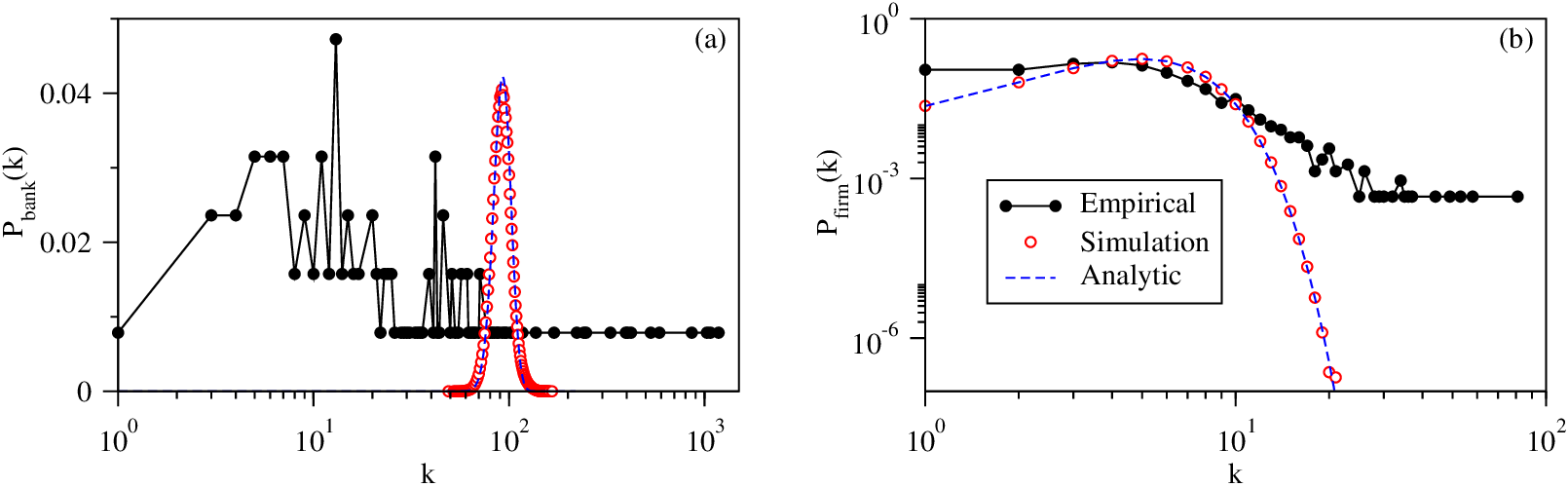}
\caption{The degree distributions $P(k)$ are plotted against degree $k$ for (a) banks and (b) firms.
Empirical, simulated and analytic results are indicated with different legends. 
} 
\label{fig2}
\end{figure}

Fig~\ref{fig1} shows the connectance $p=\langle E \rangle/MN$ as a function of $\theta$ for the model. 
The results indicate an excellent match between the analytic solution and simulation results. Simulation
results are obtained using Markov chain Monte Carlo method as explained in section~\ref{Method}. The data points 
are averaged over $1000$ independent runs. The maximum standard deviation in the data points is found to be $\sigma=0.001$. 
For the Japanese
bipartite network of banks and firms the analytic result gives $\theta = 3.1167$ and our simulation estimates 
$\theta = 3.1166 \pm 0.0015$ reflecting the sparse nature of the network.

The degree distribution $p(k)$ implies the number of nodes with degree $k$ of this model has binomial form. 
For the bank firm network, the degree distribution of the banks can be written as 
$P_{bank}(k) = \binom{N}{k}p^k(1-p)^{(N-k)}$ and for firms $P_{firm}(k) = \binom{M}{k}p^k(1-p)^{(M-k)}$.
As can be seen from the Fig.~\ref{fig2}, the degree distribution of the model does not fit with the empirical 
distribution which has a much broader shape for both the bank and firm. We conclude that the Bernoulli model is a poor model for the Japanese
bipartite network of banks and firms.

\subsection{Two star model of a bipartite network}
The two star model is an ERG model where the expected values for the total number of links and total number of two star (.i.e. path length 2) are 
constant.
The Hamiltonian for the model of a bipartite network can be written as 
$$H (x)= \theta_L Z_L(x) + \theta_{2SB} Z_{2SB}(x) + \theta_{2SF} Z_{2SF}(x)$$

Where, the total number of links $$Z_L(x) = \sum_{i=1}^{N}  \sum_{j=1}^{M} A_{ij}$$

The total number of bank two star $$Z_{2SB} =  \frac{1}{2}\sum_{i=1}^{M}  \sum_{j,k=1}^{N} (1-\delta_{jk})A_{ij}A_{ik}
= \frac{1}{2}\sum_{i=1}^{M}  \sum_{j,k=1}^{N} A_{ij}A_{ik}-\frac{1}{2}\sum_{i=1}^{M}  \sum_{j=1}^{N} A_{ij}$$
The total number of firm two star
$$Z_{2SF} =  \frac{1}{2}\sum_{i=1}^{N}  \sum_{j,k=1}^{M} (1-\delta_{jk})A_{ij}A_{ik}
= \frac{1}{2}\sum_{i=1}^{N}  \sum_{j,k=1}^{M} A_{ij}A_{ik}-\frac{1}{2}\sum_{i=1}^{N}  \sum_{j=1}^{M} A_{ij}$$ 

$\theta$'s are the associated parameters to the network observables.

The Hamiltonian $H$ can be written in terms of the adjacency matrix as follows:

 $$H  = \frac{1}{2}\sum_{i=1}^{N}  \sum_{j=1}^{M} A_{ij}(2\theta_L-\theta_{2SB}- \theta_{2SF}+ \theta_{2SB} \sum\limits_{k=1}^N A_{ik} + \theta_{2SF}\sum\limits_{k=1}^M A_{ik})$$
 Using mean-field technique of statistical physics we can set the average connection probability between any two nodes
is $p = \langle A_{ik} \rangle = A_{ik}$ by ignoring the local fluctuations. 
\begin{align*} 
      H & = \frac{1}{2}\sum_{i=1}^{N}  \sum_{j=1}^{M} A_{ij}(2\theta_L-\theta_{2SB}- \theta_{2SF} + \theta_{2SB} \sum\limits_{k=1}^N \langle A_{ik} \rangle+\theta_{2SF}\sum\limits_{k=1}^M \langle A_{ik} \rangle)\\
          & = \frac{1}{2}\sum_{i=1}^{N}  \sum_{j=1}^{M} A_{ij}(2\theta_L-\theta_{2SB}- \theta_{2SF}+ \theta_{2SB} N p +\theta_{2SF} M p) \\
          & = \Theta \sum_{i=1}^{N}  \sum_{j=1}^{M} A_{ij}
\end{align*}

Where we define $\Theta = \frac{1}{2} (2\theta_L-\theta_{2SB}- \theta_{2SF}+ \theta_{2SB} N p +\theta_{2SF} M p)$

As the Hamiltonian becomes linear with $A_{ij}$, we can easily calculate the partition function $\kappa = [1+exp(\Theta)]^{NM}$

From the partition function we can calculate other network observables:

Free energy $F=ln(\kappa)=NM \ln[1+exp(\Theta)]$

Total expected number of links
 $<Z_L> = \pdv{F}{\theta_L} = NM \frac{exp(\Theta)}{1+exp(\Theta)}$
 
This gives 
 \begin{align*} 
 p & = \frac{<Z_L>}{NM}\\
   & =\frac{exp(\Theta)}{1+exp(\Theta)}\\
   & = \frac{1}{2}[1+\tanh(\Theta/2)]\\
   & =\frac{1}{2}[1+\tanh\{0.25(2\theta_L-\theta_{2SB}- \theta_{2SF}+ \theta_{2SB} N p +\theta_{2SF} M p)\}]
\end{align*}
For convince let us define $B = 0.25(2\theta_L-\theta_{2SB}- \theta_{2SF})$ and $2J = 0.25(\theta_{2SB} N  +\theta_{2SF} M )$.

It gives, 
$$p=\frac{1}{2}[1+\tanh(B+2Jp)]$$

The solution of this transcendental equation is well known~\cite{park2004statistical}.
It has only one solution if $J\leq 1$, but if $J > 1$ it may have either one solution or three solutions (where outer two are stable solution). 
It can be shown~\cite{coolen2017generating} that the three solutions appears when $B_+(J) < B < B_-(J)$, where 
$$ B_{+}(J) = \frac{1}{2}\log[\frac{\sqrt{J}+\sqrt{J-1}}{\sqrt{J}-\sqrt{J-1}}]-\frac{\sqrt{J}}{\sqrt{J}-\sqrt{J-1}}$$

and
$$ B_{-}(J) = \frac{1}{2}\log[\frac{\sqrt{J}-\sqrt{J-1}}{\sqrt{J}+\sqrt{J-1}}]-\frac{\sqrt{J}}{\sqrt{J}+\sqrt{J-1}}$$

\begin{figure}
\includegraphics[width=\textwidth]{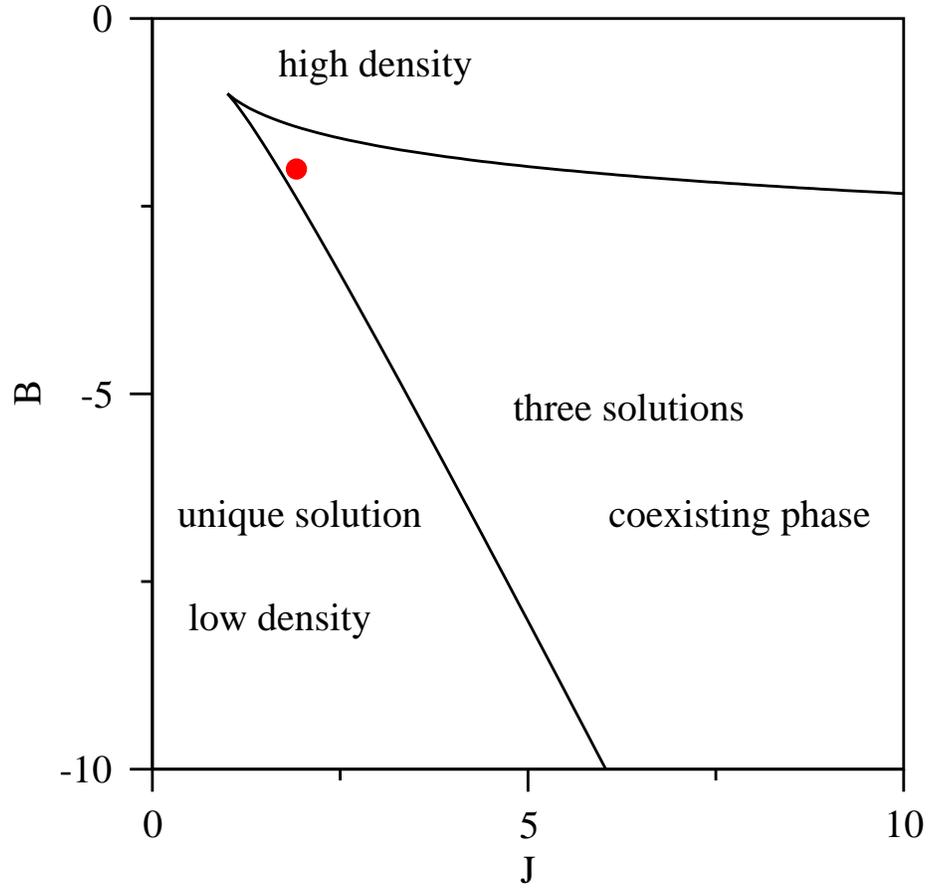}
\caption{The phase diagram for the two star model. The red circle indicates the position for the Japanese bipartite network of banks and firms for the year 2005.} 
\label{fig3}
\end{figure}

\begin{table}
\caption{Estimated values of the coupling parameters of the two star model for the Japanese bipartite network for the year 2005.}\label{tab1}
\begin{tabular}{|l|l|l|}
\hline
Parameters &  Estimated values & Standard deviation \\
\hline
$\theta_L$ &  $-3.974$ & $4.040 \times 10^{-3}$ \\
$\theta_{2SF}$&  $6.307 \times 10^{-2}$ & $3.445\times 10^{-4}$ \\
$\theta_{2SB}$ & $ 3.334  \times 10^{-3}$ & $1.328  \times 10^{-6}$\\
\hline
\end{tabular}
\end{table}
We show the phase diagram of the model in $(B-J)$ plane in Fig.~\ref{fig3}. It has three distinct regions - high density, low density and co-existence phase. 
Our estimates of the parameters are given in Table~\ref{tab1}. The values of the estimated parameters give $B = 2.004$ and $J=1.917$. As can be seen from Fig.~\ref{fig3}
at these parameter values, the system can show high fluctuation in behaviours having two coexisting phase. We conclude that the Japanese bipartite network of banks and firms 
are close to the transition point which indicate a fragile nature of the system. 

\begin{figure}
\includegraphics[width=\textwidth]{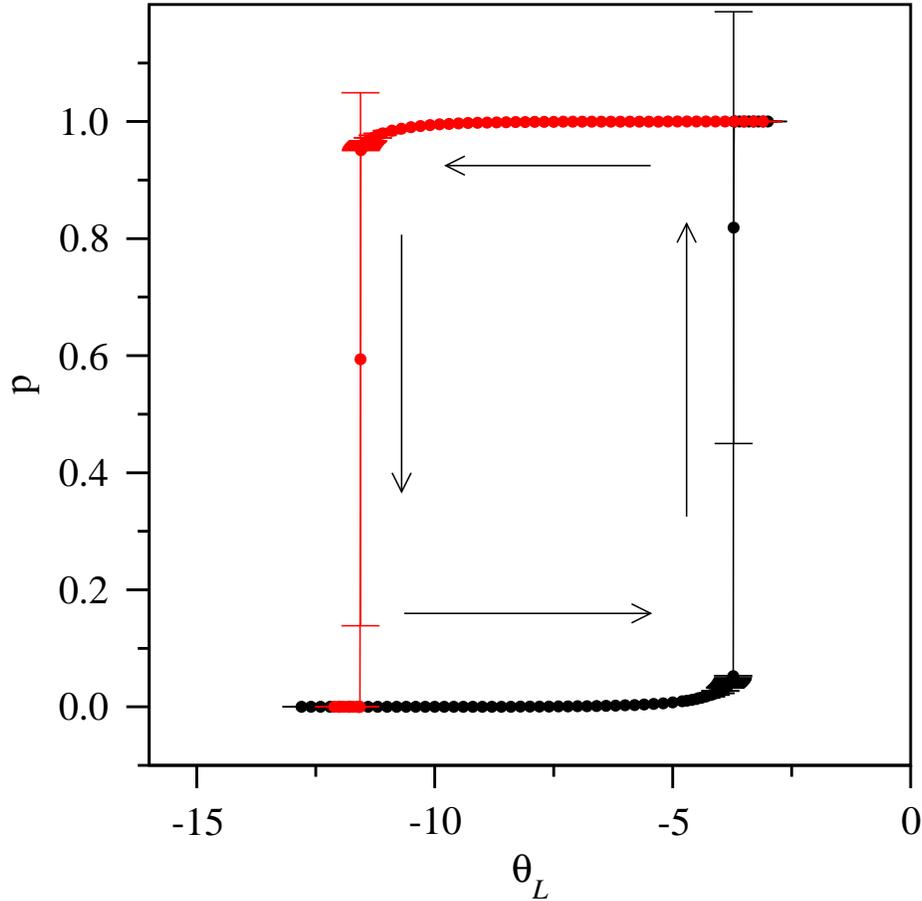}
\caption{The hysteresis plot for the two star model of the bank-firm network.
The black curve indicates the variation of connectance $p$ when $\theta_L$ increases from low to high and red curve indicates when $\theta_L$ decreases from high to low.
The values of $\theta_{2SF}$ and $\theta_{2SB}$ are kept constant as in Table~\ref{tab1}. The error bars indicate standard deviation in $p$.  
} \label{fig4}
\end{figure}

This model exhibits hysteresis behaviour as shown in Fig~\ref{fig4}. The finite area within the loop is a signature of a discontinuous transition (.i.e. first order). The first order transition 
is very dangerous for an economic network. It indicates the network can collapse suddenly if there is a slight change in the parameter values.

\begin{figure}
\includegraphics[width=\textwidth]{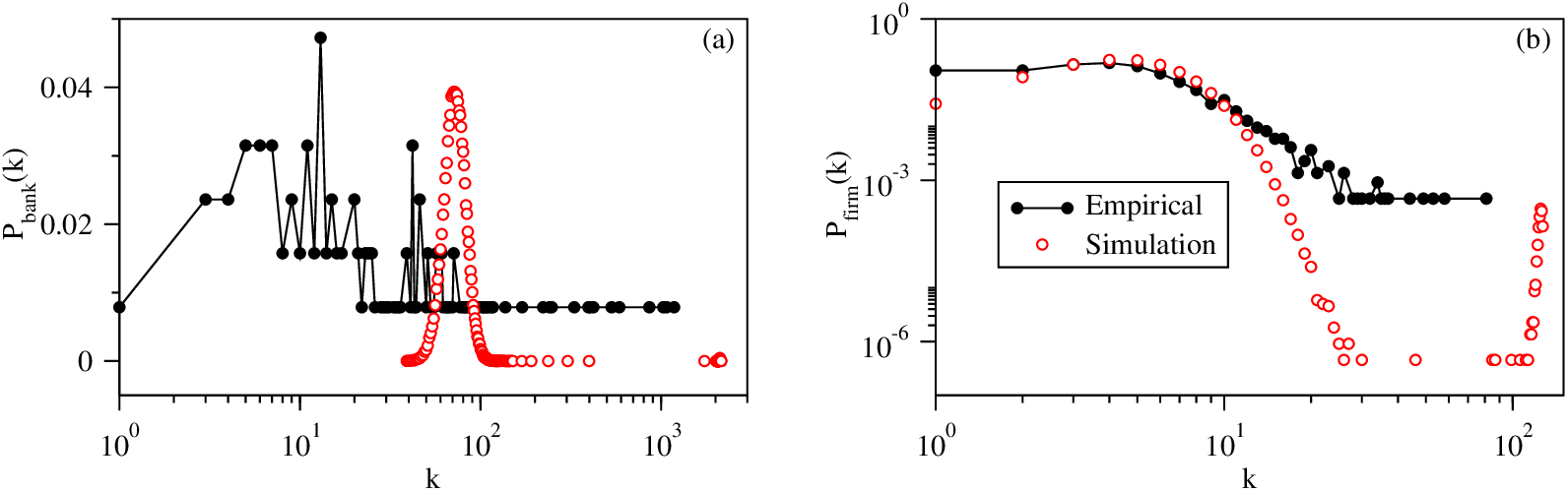}
\caption{The degree distributions $P(k)$ are plotted against degree $k$ for (a) banks and (b) firms.
Empirical and simulated results are indicated with different legends. } \label{fig5}
\end{figure}

Fig.~\ref{fig5} shows the degree distribution for the two star model. The distribution has a bi-modal nature~\cite{coolen2017generating}. 
Although the second peak near $k=N$ for the degree distribution of banks is very small. This model also can not replicate the empirical
nature of the degree distribution. In the future, we will consider more complex Hamiltonians that include endogenous as well as exogenous parameters to describe the system
in much better way. 

\section{Conclusions}
We have studied the Japanese bipartite network of banks and firms using the Bernoulli model and the two star model. 
The Bernoulli model assumes that links are formed between banks and firms independently. However, this model does not fit well
with the empirical network structure indicating a relationship present between the network structure and some hidden variables. 
As a first approximation, we consider two star model that assumes adjacent links play role in the link formation. This model
indicates that the Japanese bipartite network of banks and firms has a fragile nature. Although this model also can not capture
the empirical network structure fully.

In the future, we would like to consider the more complex Hamiltonians as well as the temporal evolution of the system in the phase space.
We believe such complex Hamiltonians will be useful to understand the network structure in detail.

%
%
%
 \bibliographystyle{splncs}
 \bibliography{ERGM}
%




\end{document}